\documentclass[12pt]{article}
\usepackage{amssymb,amsmath,epsfig}
\allowdisplaybreaks
\begin{document}

\title{\bf Lorentz Distributed Noncommutative $F(T,T_G)$ Wormhole Solutions}

\author{M. Sharif \thanks{msharif.math@pu.edu.pk} and
Kanwal Nazir \thanks{awankanwal@yahoo.com}~~\thanks{On leave from
Department of Mathematics, Lahore College
for Women University, Lahore-54000, Pakistan.}\\
Department of Mathematics, University of the Punjab,\\
Quaid-e-Azam Campus, Lahore-54590, Pakistan.}

\date{}
\maketitle
\begin{abstract}
The aim of this paper is to study static spherically symmetric
noncommutative $F(T,T_G)$ wormhole solutions along with Lorentzian
distribution. Here, $T$ and $T_G$ are torsion scalar and
teleparallel equivalent Gauss-Bonnet term, respectively. We take a
particular redshift function and two $F(T,T_G)$ models. We analyze
the behavior of shape function and also examine null as well as weak
energy conditions graphically. It is concluded that there exist
realistic wormhole solutions for both models. We also study the
stability of these wormhole solutions through equilibrium condition
and found them stable.
\end{abstract}

\section{Introduction}

This is well-known through different cosmological observations that
our universe undergoes accelerated expansion that opens up new
directions. A plethora of work has been performed to explain this
phenomenon. It is believed that behind this expansion, there is a
mysterious force dubbed as dark energy (DE) identified by its
negative pressure. Its nature is generally described by the
following two well-known approaches. The first approach leads to
modify the matter part of general relativity (GR) action that gives
rise to several DE models including cosmological constant,
k-essence, Chaplygin gas, quintessence etc \cite{1}-\cite{5}.

The second way leads to gravitational modification which results in
modified theories of gravity. Among these theories, the $F(T)$
theory \cite{8} is a viable modification which is achieved by
torsional formulation. Various cosmological features of this theory
have been investigated such as solar system constraints, static
wormhole solutions, discussion of Birkhoff's theorem, instability
ranges of collapsing stars and many more \cite{11}-\cite{11b}.
Recently, a well known modified version of $F(T)$ theory is proposed
by involving higher order torsion correction terms named as
$F(T,T_G)$ theory depend upon $T$ and $T_G$ \cite{16a}. This is a
completely different theory which does not correspond to $F(T)$ as
well as any other modified theory. It is a novel modified gravity
theory having no curvature terms. The dynamical analysis \cite{16}
and cosmological applications \cite{16b} of this theory turn out to
be very captivating.

Chattopadhyay et al. \cite{18} studied pilgrim DE model and
reconstructed $F(T,T_{G})$ models by assuming flat FRW metric. Jawad
et al. \cite{19} explored reconstruction scheme in this theory by
considering a particular ghost DE model. Jawad and Debnath \cite{17}
worked on reconstruction scenario by taking a new pilgrim DE model
and evaluated different cosmological parameters. Zubair and Jawad
discussed thermodynamics at the apparent horizon \cite{17a}. We
developed reconstructed models by assuming different eras of DE and
their combinations with FRW and Bianchi type I universe models,
respectively \cite{20}.

The study of wormhole solutions provides fascinating aspects of
cosmology especially in modified theories. Agnese and Camera
\cite{24a} discussed static spherically symmetric and traversable
wormhole solutions in Brans-Dicke scalar tensor theory. Anchordoqui
et al. \cite{25a} showed the existence of analytical wormhole
solutions and concluded that there may exist a wormhole sustained by
normal matter. Lobo and Oliveira \cite{21} considered $f(R)$ theory
to examine the traversable wormhole geometries through different
equations of state. They analyzed that wormhole solution may exist
in this theory and discussed the behavior of energy conditions.
B$\ddot{o}$hmer et al. \cite{22} examined static traversable $F(T)$
wormhole geometry by considering a particular $F(T)$ model and
constructed physically viable wormhole solutions. The dynamical
wormhole solutions have also been studied in this theory by assuming
anisotropic fluid \cite{23}. Recently, Sharif and Ikram \cite{24}
explored static wormhole solutions and investigated energy
conditions in $f(G)$ gravity. They found that these conditions are
satisfied only for barotropic fluid in some particular regions.

General relativity does not explain microscopic physics (completely
described through quantum theory). Classically, the smooth texture
of spacetime damages at short distances. In GR, the spacetime
geometry is deformed by gravity while it is quantized through
quantum gravity. To overcome this problem, noncommutative geometry
establishes a remarkable framework that discusses the dynamics of
spacetime at short distances. This framework introduces a scale of
minimum length having a good agreement with Planck length. The
consequences of noncommutativity can be examined in GR by taking the
standard form of the Einstein tensor and altered form of matter
tensor.

Noncommutative geometry is considered as the essential property of
spacetime geometry which plays an impressive role in several areas.
Rahaman et al. \cite{29} explored wormhole solutions along with
noncommutative geometry and showed the existence of asymptotically
flat solutions for four dimensions. Abreu and Sasaki \cite{30}
studied the effects of null (NEC) and weak (WEC) energy conditions
with noncommutative wormhole. Jamil et al. \cite{31} discussed the
same work in $f(R)$ theory. Sharif and Rani \cite{32} investigated
wormhole solutions with the effects of electrostatic field and for
galactic halo regions in $F(T)$ gravity.

Recently, Bhar and Rahaman \cite{33} considered Lorentzian
distributed density function and examined that wormhole solutions
exist in different dimensional spacetimes with noncommutative
geometry. They found that wormhole solutions can exist only in four
and five dimensions but no wormhole solution exists for higher than
five dimension. Jawad and Rani \cite{34} investigated Lorentz
distributed noncommutative wormhole solutions in $F(T)$ gravity. We
have explored noncommutative geometry in $F(T,T_G)$ gravity and
found that effective energy-momentum tensor is responsible for the
violation of energy conditions rather than noncommutative geometry
\cite{35}. Inspired by all these attempts, we investigate whether
physically acceptable wormholes exist in $F(T,T_G)$ gravity along
with noncommutative Lorentz distributed geometry. We study wormhole
geometry and corresponding energy conditions.

The paper is arranged as follows. Section \textbf{2} briefly recalls
the basics of $F(T,T_G)$ theory, the wormhole geometry and energy
conditions. In section \textbf{3}, we investigate physically
acceptable wormhole solutions and energy conditions for two
particular $F(T,T_G)$ models. In section \textbf{4}, we analyze the
stability of these wormhole solutions. Last section summarizes the
results.

\section{$F(T,T_G)$ Gravity}

This section presents some basic review of $F(T,T_G)$ gravity. The
idea of such extension is to construct an action involving higher
order torsion terms. In curvature theory other than simple
modification as $f(R)$ theory, one can propose the higher order
curvature correction terms in order to modify the action such as GB
combination $G$ or functions $f(G)$. In a similar way, one can start
from the teleparallel theory and construct an action by proposing
higher torsion correction terms.

The most dominant variable in the underlying gravity is the tetrad
field $e_{a}(x^{\lambda})$. The simplest one is the trivial tetrad
which can be expressed as
$e_{a}={\delta^{\lambda}}_{a}\partial_{\lambda}$ and
$e^{b}={\delta_{\lambda}}^{b}\partial^{\lambda}$, where the
Kronecker delta is denoted by $\delta^{\lambda}_{a}$. These tetrad
fields are of less interest as they result in zero torsion. On the
other hand, the non-trivial tetrad field is more favorable to
construct teleparallel theory because they give non-zero torsion.
They can be expressed as
\begin{equation}\nonumber
h_{a}={h_{a}^{~\lambda}}\partial_{\lambda},\quad
h^{b}={h^{a}_{~\lambda}}dx^{\lambda}.
\end{equation}
The non-trivial tetrad satisfies
$h_{~\lambda}^{a}h^{~\lambda}_{b}=\delta^{a}_{b}$ and
$h_{~\lambda}^{a}h^{~\mu}_{a}=\delta^{\mu}_{\lambda}$. The tetrad
fields can be related with metric tensor through
\begin{equation}\nonumber
g_{\lambda\mu}=\eta_{ab}h_{\lambda}^{a}h^{b}_{\mu},
\end{equation}
where $\eta_{ab}=diag(1,-1,-1,-1)$ is the Minkowski metric. Here,
Greek indices $(\lambda,\mu)$ represent coordinates on manifold and
Latin indices $(a, b)$ correspond to the coordinates on tangent
space. The other field is described as the connection 1-forms
${\omega^{a}}_{b}(x^{\lambda})$ which are the source of parallel
transportation, also known as Weitzenb$\ddot{o}$ck connection. It
has the following form
\begin{equation}\nonumber
\omega^{\mu}_{\lambda\nu}={h^{\mu}}_{a}{h^{a}}_{\lambda,\nu}.
\end{equation}

The structure coefficients $C^{c}_{ab}$ appear in commutation
relation of the tetrad as
\begin{equation}\nonumber
C^{c}_{ab}=h_{c}^{-1}[h_{a},h_{b}],
\end{equation}
where
\begin{equation}\nonumber
C^{c}_{ab}={h^{\mu}}_{b}{h^{\lambda}}_{a}({h^{c}}_{\lambda,\mu}-{h^{c}}_{\mu,\lambda}).
\end{equation}
The torsion as well as curvature tensors has the following
expressions
\begin{eqnarray}\nonumber
T^{a}_{bc}&=&-C^{a}_{bc}-\omega^{a}_{bc}+\omega^{a}_{cb},\\\nonumber
R^{a}_{bcd}&=&\omega^{e}_{bd}\omega^{a}
_{ec}+\omega^{a}_{bd,c}-C^{e}_{cd}\omega^{a}_{be}
-\omega^{e}_{bc}\omega^{a}_{ed}-\omega^{a}_{bc,d}.
\end{eqnarray}
The contorsion tensor can be described as
\begin{equation}\nonumber
K_{abc}=-K_{bac}=\frac{1}{2}(-T_{abc}+T_{cab}-T_{bca}).
\end{equation}
Both the torsion scalars are written as
\begin{eqnarray}\nonumber
T&=&\frac{1}{4}T^{abc}T_{abc}+\frac{1}{2}T^{abc}T_{cba}
-T_{ab}^{~~a}T^{cb}_{~~c},\\\nonumber
T_G&=&({K^{ea_{2}}}_{b}{K^{a_{3}}}_{fc}{K^{a_{1}}}_{ea}
{K^{fa_{4}}}_{d}+2{K^{ea_{4}}}_{f}{K^{f}}_{cd}
{K^{a_{1}a_{2}}}_{a}{{K^{a_{3}}}_{eb}}
+2{K^{ea_{4}}}_{c,d}{K^{a_{3}}}_{eb}\\\nonumber
&\times&{K^{a_{1}a_{2}}}_{a}-2{{K^{a_{3}}}_{eb}{K^{e}}
_{fc}K^{a_{1}a_{2}}}_{a}{K^{fa_{4}}}_{d})
\delta^{abcd}_{a_{1}a_{2}a_{3}a_{4}}.
\end{eqnarray}
This comprehensive theory has been proposed by Kofinas and Saridakis
\cite{16} whose action is described as
\begin{equation}\nonumber
S=\int
h\left[\frac{F(T,T_G)}{\kappa^{2}}+\mathcal{L}_{m}\right]d^4x,
\end{equation}
where $\mathcal{L}_{m}$ is the matter Lagrangian, $\kappa^{2}=1$,
$g$ represents determinant of the metric coefficients and
$h=\sqrt{-g}=\det(h^a_\lambda)$. The field equations obtained by
varying the action about $h^a_\lambda$ are given as
\begin{eqnarray}\nonumber
&&C^{b}_{~cd}(H^{dca}+2H^{[ac]d})+(-T_G F_{T_G}(T,T_G)+F(T,T_G)-T
F_{T}(T,T_G))\eta^{ab}\\\nonumber
&+&2(H^{[ba]c}-H^{[kcb]a}+H^{[ac]b})C_{~dc}^{d}+2(-H^{[cb]a}
+H^{[ac]b}+H^{[ba]c})_{,c}+4H^{[db]c}\\\label{1}&\times&
C_{(dc)}^{~~~a}+T^{a}_{~cd}H^{cdb}-\mathcal{H}^{ab}=\kappa^2\mathcal{T}^{ab},
\end{eqnarray}
where
\begin{eqnarray}\nonumber
H^{abc}&=&(\eta^{ac}K^{bd}_{~~d}-K^{bca})F_{T}(T,T_{G})+F_{T_{G}}(T,T_{G})[
(\epsilon^{ab}_{~~lf}K^{d}_{~qr}K^{l}_{~dp}\\\nonumber
&+&2K^{bc}_{~~p}\epsilon^{a}_{~dlf}K^{d}_{~qr}
+K^{il}_{~~p}\epsilon_{qdlf}K^{jd}_{~~r})K^{qf}_{~~t}\epsilon^{kprt}
+\epsilon^{ab}_{~~ld}K^{fd}_{~~p}\epsilon^{cprt}(K^{l}_{~fr,t}
\\\nonumber&-&\frac{1}{2}C^{q}_{~tr}K^{l}_{~fq})
+\epsilon^{cprt}K^{df}_{~p}\epsilon^{al}_{~~df}(K^{b}_{~kr,t}
-\frac{1}{2}C^{q}_{~tr}K^{b}_{~lq})]
+\epsilon^{cprt}\epsilon^{a}_{~ldf}\\\nonumber&\times&[F_{T_{G}}(T,T_{G})
K^{bl}_{~~[q}K^{df}_{~~r]}C^{q}_{~pt}+(K^{bl}_{~p}F_{T_{G}}(T,T_{G})
K^{df}_{~r})_{,t}],\\\nonumber
\mathcal{H}^{ab}&=&F_{T}(T,T_G)\epsilon^{a}_{~lce}K^{l}
_{~fr}\epsilon^{brte}K^{fc}_{~~t}.
\end{eqnarray}
Here, $\mathcal{T}^{ab}$ represents the matter energy-momentum
tensor. The functions $F_{T}$ and $F_{T_{G}}$ are the derivatives of
$F$ with respect to $T$ and $T_{G}$, respectively. Notice that for
$F(T,T_{G})=-T$, teleparallel equivalent to GR is achieved. Also,
for $T_G=0$, we can obtain $F(T)$ theory.

Next, we explain the wormhole geometry as well as energy conditions
in this gravity.

\subsection{Wormhole Geometry}

Wormholes associate two disconnected models of the universe or two
distant regions of the same universe (interuniverse or intrauniverse
wormhole). It has basically a tube, bridge or tunnel type
appearance. This tunnel provides a shortcut between two distant
cosmic regions. The well-known example of such a structure is
defined by Misner and Wheeler \cite{36} in the form of solutions of
the Einstein field equations named as wormhole solutions. Einstein
and Rosen made another attempt and established Einstein-Rosen
bridge.

The first attempt to introduce the notion of traversable wormholes
is made by Morris and Thorne \cite{37}. The Lorentzian traversable
wormholes are more fascinating in a way that one may traverse from
one to another end of the wormhole \cite{37a}. The traversability is
possible in the presence of exotic matter as it produces repulsion
which keeps open throat of the wormhole. Being the generalization of
Schwarzschild wormhole, these wormholes have no event horizon and
allow two way travel. The spacetime for static spherically symmetric
as well as traversable wormholes is defined as \cite{37}
\begin{equation}\label{2}
ds^2=e^{2\alpha(r)}dt^2-\frac{dr^2}{\left(1-\frac{\beta(r)}{r}\right)}
-r^2d\theta^2-r^2\sin^2\theta d\phi^2,
\end{equation}
where $\alpha(r)$ is the redshift function and $\beta(r)$ represents
the shape function. The gravitational redshift is measured through
the function $\alpha(r)$ whereas $\beta(r)$ controls the wormhole
shape. The radial coordinate $r$, redshift and shape functions must
satisfy few conditions for the traversable wormhole. The redshift
function needs to satisfy no horizon condition because it is
necessary for traversability. Thus to avoid horizons, $\alpha(r)$
must be finite throughout. For this purpose, we assumed zero
redshift function that implies $e^{2\alpha(r)}\rightarrow1$. There
are two properties related to the shape function to maintain the
wormhole geometry. The first property is positiveness, i.e., as
$r\rightarrow\infty$, $\beta(r)$ must be defined as a positive
function. The second is flaring-out condition, i.e.,
$\left(\frac{\beta(r)-r\beta'(r)}{\beta^2(r)}\right)>0$ and
$\beta(r)=r_{th}$ at $r=r_{th}$ with $\beta'(r_{th})<1$ ( $r_{th}$
is the wormhole throat radius). The condition of asymptotic flatness
($\frac{\beta(r)}{r}\rightarrow0$ as $r\rightarrow\infty$) should be
fulfilled by the spacetime at large distances.

To investigate the wormhole solutions, we assume a diagonal tetrad
\cite{37} as
\begin{eqnarray}\label{3}
h^{a}_{\lambda}=diag\left(e^{-\alpha(r)},\frac{1}{\left(1-\frac{\beta(r)}{r}\right)}
,~r,~r\sin\theta\right).
\end{eqnarray}
This is the simplest and frequently used tetrad for the Morris and
Thorne static spherically symmetric metric. This also provides
non-zero $T_\mathcal{G}$ which is the basic ingredient for this
theory. If we take some other tetrad then it may lead to zero
$T_\mathcal{G}$. Thus these orthonormal basis are most suitable for
this theory. The torsion scalars turn out to be
\begin{eqnarray}\label{4}
T&=&\frac{4}{r}\left(1-\frac{\beta(r)}{r}\right)\alpha'+\frac{2}{r^2}\left(1-\frac{\beta(r)}{r}\right),\\\nonumber
T_G&=&\frac{8\beta(r)\alpha'(r)}{r^4}-\frac{8\beta(r)\alpha'^2(r)}{r^3}
\left(1-\frac{\beta(r)}{r}\right)+\frac{12\beta(r)\alpha'(r)\beta'(r)}{r^4}\\\label{5}
&-&\frac{8\beta'(r)\alpha'(r)}{r^3}
-\frac{12\beta^2(r)\alpha'(r)}{r^5}-\frac{8\beta(r)\alpha''(r)}{r^3}
\left(1-\frac{\beta(r)}{r}\right).
\end{eqnarray}
In order to satisfy the condition of no horizon for a traversable
wormhole, we have to assume $\alpha(r)=0$. Substituting this
assumption in the above torsion scalars, we obtain $T_G=0$ which
means that the function $F(T,T_G)$ reduces to $F(T)$ representing
$F(T)$ theory. Hence, we cannot take $\alpha(r)$ as a constant
function instead we assume $\alpha(r)$ as
\begin{eqnarray}\label{6}
\alpha(r)=-\frac{\psi}{r},\quad\psi>0,
\end{eqnarray}
which is finite and non-zero for $r>0$. Also, it satisfies
asymptotic flatness as well as no horizon condition. We assume that
anisotropic matter threads the wormhole for which the
energy-momentum tensor is defined as
\begin{eqnarray}\nonumber
\mathcal{T}^{(m)}_{~~\lambda\mu}=(p_t+\rho)V_\lambda
V_\mu-g_{\lambda\mu}p_t+(p_r-p_t)\eta_\mu\eta_\lambda,
\end{eqnarray}
where $\rho$, $V_\lambda$, $\eta_\lambda$, $p_r$ and $p_t$ represent
the energy density, four-velocity, radial spacelike four-vector
orthogonal to $V_\lambda$, radial and tangential components of
pressure, respectively. We consider energy-momentum tensor as
$\mathcal{T}^{(m)}_{~~\lambda\mu}=diag(\rho, -p_r,-p_t, -p_t)$.
Using Eqs.(\ref{2})-(\ref{6}) in (\ref{1}), we obtain the field
equations as
\begin{eqnarray}\nonumber
\rho&=&F(T,T_G)+\frac{2\beta'(r)}{r^2}F_T(T,T_G)
-T_GF_{T_G}(T,T_G)-TF_T(T,T_G)-\frac{4F_{T}'}{r}(1\\\nonumber&-&
\frac{\beta(r)}{r})+\frac{4F_{T_G}'(T,T_G)}{r^3}\left(\frac{5\beta(r)}{r}
-\frac{3\beta^2(r)}{r^2}-2-3\beta'(r)\left(1-
\frac{\beta(r)}{r}\right)\right)\\\label{7}&+&
\frac{8F_{T_G}''(T,T_G)}{r^2}\left(1-\left(2-\frac{\beta(r)}{r}\right)
\frac{\beta(r)}{r}\right)
,\\\nonumber
p_r&=&-F(T,T_G)+F_T(T,T_G)\left(T-\frac{4\psi}{r^3}-\frac{2\beta(r)}{r^3}
+\frac{4\beta(r)\psi}{r^4}\right)
+T_GF_{T_G}(T,T_G)
\\\label{8}&+&\frac{48}{r^4}\left(1-\frac{\beta(r)}{r}\right)^2
\psi F_{T_G}'(T,T_G),\\\nonumber
p_t&=&-F(T,T_G)+T_GF_{T_G}(T,T_G)+TF_T(T,T_G)+\left(\frac{\beta(r)}{r^3}-\frac{2\psi}{r^3}
-\frac{\beta'(r)}{r^2}
\right.\\\nonumber&+&\left.\frac{\beta(r)\psi}{r^4}
+\frac{2\psi^2}{r^4}+\frac{\beta'(r)\psi}{r^3}-\frac{2\beta(r)\psi^2}{r^5}
-\frac{4\beta(r)\psi}{r^4}+\frac{4\psi}{r^3}\right)F_{T}(T,T_G)\\\nonumber
&+&2\left(\frac{1}{r}-\left(1-\frac{\beta(r)}{r}\right)\psi'+\frac{\beta(r)
\psi}{r^3}\right)F_{T}'(T,T_G)
+\left(\frac{12\psi\beta(r)}{r^5}-\frac{12\psi\beta^2(r)}{r^6}
\right.\\\nonumber&+&\left.\frac{16\beta(r)\psi^2}{r^6}
+\frac{12\psi\beta'(r)}{r^4} -\frac{8\psi^2}{r^5}
-\frac{8\beta^2(r)\psi^2}{r^7}+\frac{12\beta(r)\psi\beta'(r)}{r^5}
-\frac{16\psi}{r^4}\right.\\\nonumber&-&\left.\frac{16\beta^2(r)\psi}{r^6}-\frac{32\beta(r)\psi}{r^5}
\right)F_{T_G}'(T,T_G)+\frac{8\psi}{r^3}\left(1
-\frac{\beta(r)}{r}\left(2+\frac{\beta(r)}{r}\right)\right)\\\label{9}&\times&F_{T_G}''(T,T_G),
\end{eqnarray}
where prime stands for the derivative with respect to $r$.

\subsection{Energy Conditions}

These conditions are mostly considered in GR and also in modified
theories of gravity. As these conditions violates in GR and this
guarantees the presence of realistic wormhole. The origin of these
conditions is the Raychaudhuri equations along with the requirement
of attractive gravity \cite{38}. Consider timelike and null vector
field congruences as $u^\lambda$ and $k^\lambda$, respectively, the
Raychaudhuri equations are formulated as follows
\begin{eqnarray}\nonumber
\frac{d\Theta}{d\tau}-\omega_{\lambda\mu}\omega^{\lambda\mu}+R_{\lambda\mu}u^\lambda
u^\mu+\frac{1}{3}\Theta^2+\sigma_{\lambda\mu}\sigma^{\lambda\mu}&=&0,\\\nonumber
\frac{d\Theta}{d\chi}-\omega_{\lambda\mu}\omega^{\lambda\mu}+R_{\lambda\mu}k^\lambda
k^\mu+\frac{1}{2}\Theta^2+\sigma_{\lambda\mu}
\sigma^{\lambda\mu}&=&0,
\end{eqnarray}
where the expansion scalar $\Theta$ is used to explain expansion of
the volume and shear tensor $\sigma^{\lambda\mu}$ provides the
information about the volume distortion. The vorticity tensor
$\omega^{\lambda\mu}$ explains the rotating curves. The positive
parameters $\chi$ and $\tau$ are used to interpret the congruences
in manifold. In the above equations, we may neglect quadratic terms
as we consider small volume distortion (without rotation). Thus
these equations reduce to $\Theta=-\tau R_{\lambda\mu}u^\lambda
u^\mu=-\chi R_{\lambda\mu}k^\lambda k^\mu$. The expression
$\Theta<0$ ensures the attractiveness of gravity which leads to
$R_{\lambda\mu}u^\lambda u^\mu\geq0$ and $R_{\lambda\mu}k^\lambda
k^\mu\geq0$. In modified theories, the Ricci tensor is replaced by
the effective energy-momentum tensor, i.e.,
$\mathcal{T}_{\lambda\mu}^{(eff)}u^\lambda u^\mu\geq0$ and
$\mathcal{T}_{\lambda\mu}^{(eff)}k^\lambda k^\mu\geq0$ which
introduce effective pressure and effective energy density in these
conditions.

It is well-known that the violation of NEC is the basic ingredient
to develop a traversable wormhole (due to the existence of exotic
matter). It is noted that in GR, this type of matter leads to the
non-realistic wormhole otherwise normal matter fulfills NEC. In
modified theories, we involve effective energy density as well as
pressure by including effective energy-momentum tensor
$\mathcal{T}^{eff}_{\lambda\mu}$ in the corresponding energy
conditions. This effective energy-momentum tensor is given as
\begin{equation}\nonumber
\mathcal{T}^{(eff)}_{\lambda\mu}=\mathcal{T}^{(H)}_{\lambda\mu}+\mathcal{T}^{(m)}_{\lambda\mu},
\end{equation}
where $\mathcal{T}^{(H)}_{\lambda\mu}$ are dark source terms related
to the underlying $F(T,T_G)$ theory. The condition (violation of
NEC) related to $\mathcal{T}^{(eff)}_{\lambda\mu}$ confirms the
presence of traversable wormhole by holding its throat open. Thus,
there may be a chance for normal matter to fulfil these conditions.
Hence, there can be realistic wormhole solutions in this modified
scenario.

The four conditions (NEC, WEC, dominant (DEC) and strong energy
condition (SEC)) are described as
\begin{itemize}
\item NEC:\quad $p_n^{(eff)}+\rho^{(eff)}\geq 0$, where $n=1,2,3.$
\item WEC:\quad $p_n^{(eff)}+\rho^{(eff)}\geq0,\quad\rho^{(eff)}\geq0$,
\item DEC:\quad $p_n^{(eff)}\pm\rho^{(eff)}\geq0,\quad\rho^{(eff)}\geq0$,
\item SEC:\quad
$p_n^{(eff)}+\rho^{(eff)}\geq0,\quad\rho^{(eff)}+3p^{(eff)}\geq0$.
\end{itemize}

Solving Eqs.(\ref{7}) and (\ref{8}) for effective energy density and
pressure, we evaluate the radial effective NEC as
\begin{equation}\nonumber
p_r^{(eff)}+\rho^{(eff)}=\frac{1}{F_T(T,T_G)}\left(\frac{-\beta(r)}{r^3}
+\frac{\beta'(r)}{r^2}+\frac{2}{r}
\left(1-\frac{\beta(r)}{r}\right)\alpha'\right).
\end{equation}
So, $p_r^{(eff)}+\rho^{(eff)}<0$ represents the violation of
effective NEC as
\begin{equation}\label{9pN}
\left(\frac{-\beta(r)}{r^3}+\frac{\beta'(r)}{r^2}+\frac{2}{r}
\left(1-\frac{\beta(r)}{r}\right)\alpha'\right)<0.
\end{equation}
If this condition holds then it shows that the traversable wormhole
exists in this gravity.

\section{Wormhole Solutions}

Noncommutative geometry is the fundamental discretization of the
spacetime and it performs effectively in different areas. It plays
an important role in eliminating the divergencies that originates in
GR. In noncommutativity, smeared substances take the place of
pointlike structures. Considering the Lorentzian distribution, the
energy density of particle-like static spherically symmetric object
with mass $\mathcal{M}$ has the following form \cite{39}
\begin{equation}\label{10}
\rho_{NCL}=\frac{\mathcal{M}\sqrt{\theta}}{\pi^2(\theta+r^2)^2},
\end{equation}
where $\theta$ is the noncommutative parameter. Comparing
Eqs.(\ref{7}) and (\ref{10}), i.e., $\rho_{NCL}=\rho$, we obtain
\begin{eqnarray}\nonumber
\frac{\mathcal{M}\sqrt{\theta}}{\pi^2(\theta+r^2)^2}&=&F(T,T_G)
+\frac{2\beta'(r)F_T}{r^2}-TF_T(T,T_G)
-\frac{4F'_T(T,T_G)}{r}\left(1\right.\\\nonumber&-&\left.\frac{\beta(r)}{r}\right)-
T_GF_{T_G}(T,T_G)+\frac{4F_{T_G}'(T,T_G)}{r^3}\left(\frac{5\beta(r)}{r}
-2-\frac{3\beta^2(r)}{r^2}\right.\\\nonumber&-&\left.3\left(1-
\frac{\beta(r)}{r}\right)\beta'(r)\right)
+\frac{8F_{T_G}''(T,T_G)}{r^2}\left(1-\frac{\beta(r)}{r}
\left(2\right.\right.\\\label{11}&-&\left.\left.
\frac{\beta(r)}{r}\right)\right).
\end{eqnarray}
The above equation contains two unknown functions $F(T,T_G)$ and
$\beta(r)$. In order to solve this equation, we have to assume one
of them and evaluate the other one. Next, we consider some specific
and viable models from $F(T,T_G)$ theory and investigate the
wormhole solutions under Lorentzian distributed noncommutative
geometry. We also discuss the corresponding energy conditions.

\subsection{First Model}
\begin{figure}\center
\epsfig{file=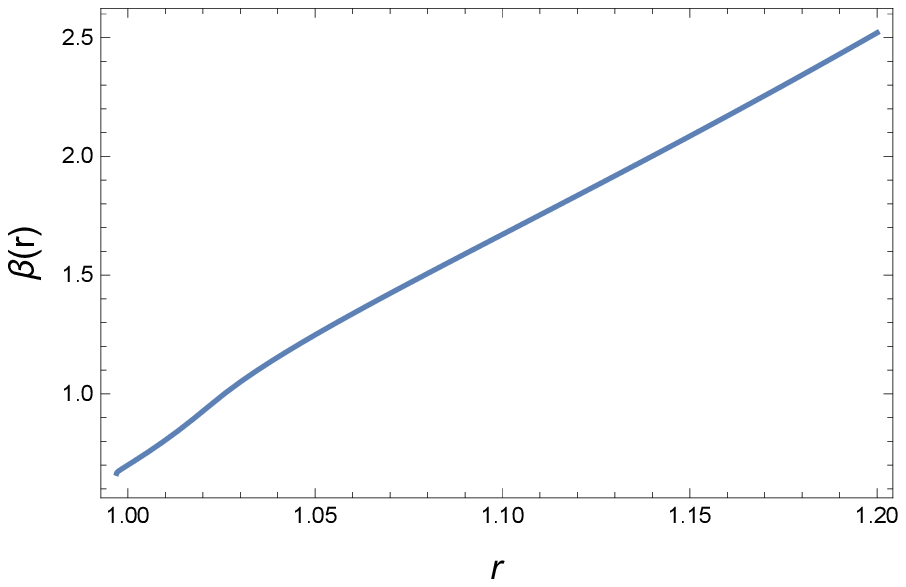,width=0.45\linewidth}
\epsfig{file=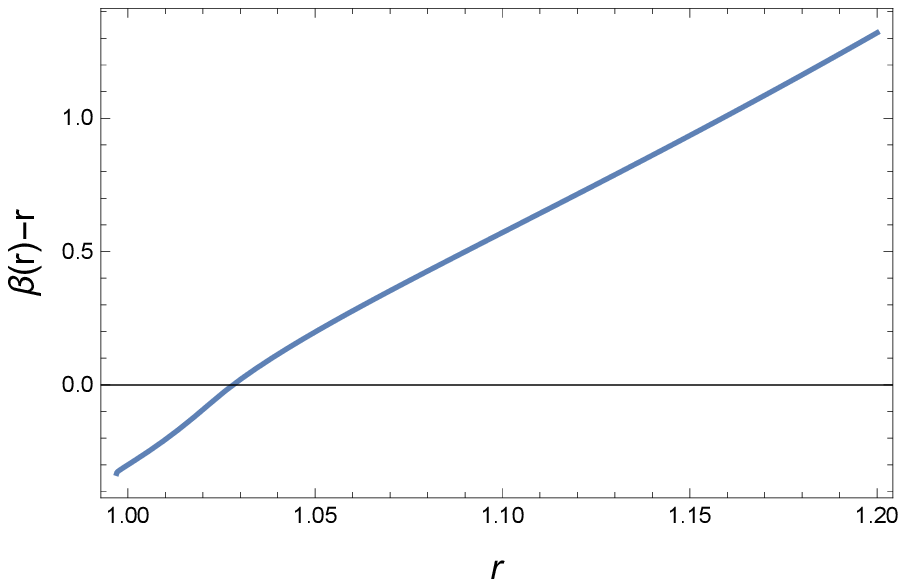,width=0.45\linewidth}
\epsfig{file=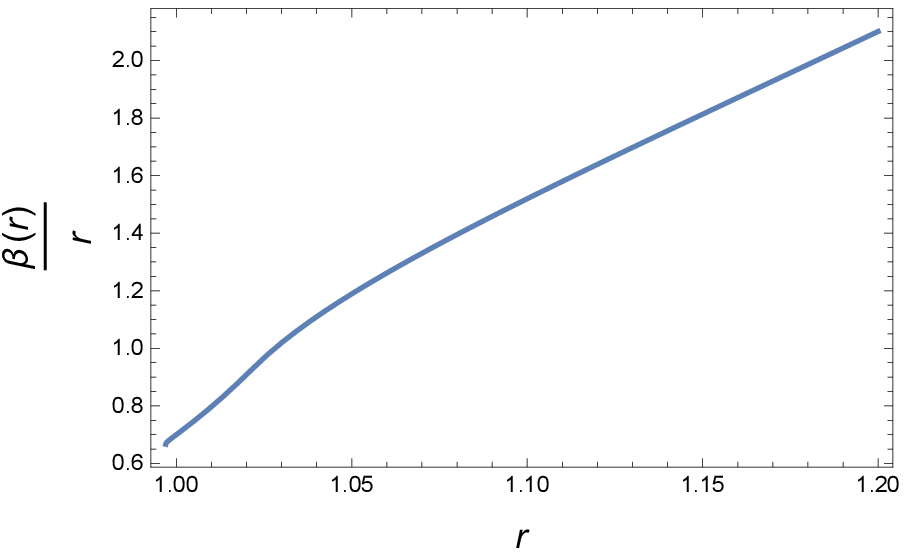,width=0.5\linewidth}\caption{Plots of $\beta(r)$,
$\beta(r)-r$ and $\frac{\beta(r)}{r}$ versus $r$ for the first
model.}
\end{figure}

The first model is considered as \cite{17a}
\begin{eqnarray}\label{12}
F(T,T_G)=-T+\gamma_1(T^2+\gamma_2T_G)+\gamma_3(T^2+\gamma_4T_G)^2,
\end{eqnarray}
where $\gamma_{1}$, $\gamma_{2}$, $\gamma_{3}$ and $\gamma_{4}$ are
arbitrary constants. Here, we take $\gamma_{2}$ and $\gamma_{4}$ as
dimensionless whereas $\gamma_{1}$ and $ \gamma_{3}$ have dimensions
of lengths. This model involves second order $T_G$ terms and fourth
order contribution from torsion term $T$. Using Eqs.(\ref{4}),
(\ref{5}) and (\ref{12}) in (12), we achieve a complicated
differential equation in terms of $\beta(r)$ that cannot be handled
analytically. So, we solve it numerically by choosing the
corresponding parameters as $\gamma_1=81$, $\gamma_2=-0.0091$,
$\gamma_3=12$, $\gamma_4=32$. The values of the remaining parameters
$\mathcal{M}=15$, $\theta=0.5$ and $\psi=1$ are taken from
\cite{34}. To  plot the graph of $\beta(r)$, we take the initial
values as $\beta(1)=0.7$, $\beta'(1)=9.9$ and $\beta''(1)=5.5$.
Figure \textbf{1} (left panel) represents the increasing behavior of
shape function $\beta(r)$. We discuss the wormhole throat by
plotting $\beta(r)-r$ in the right panel.
\begin{figure}\center
\epsfig{file=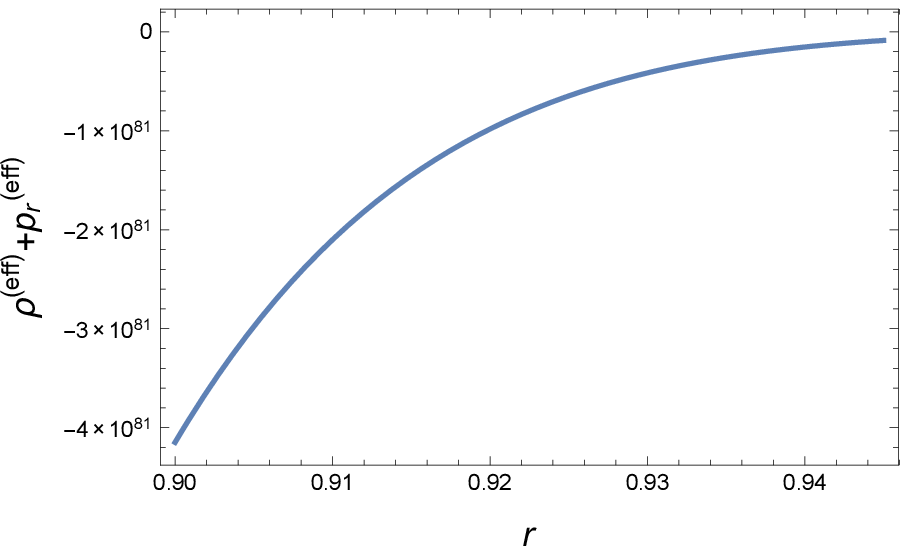,width=0.49\linewidth}
\epsfig{file=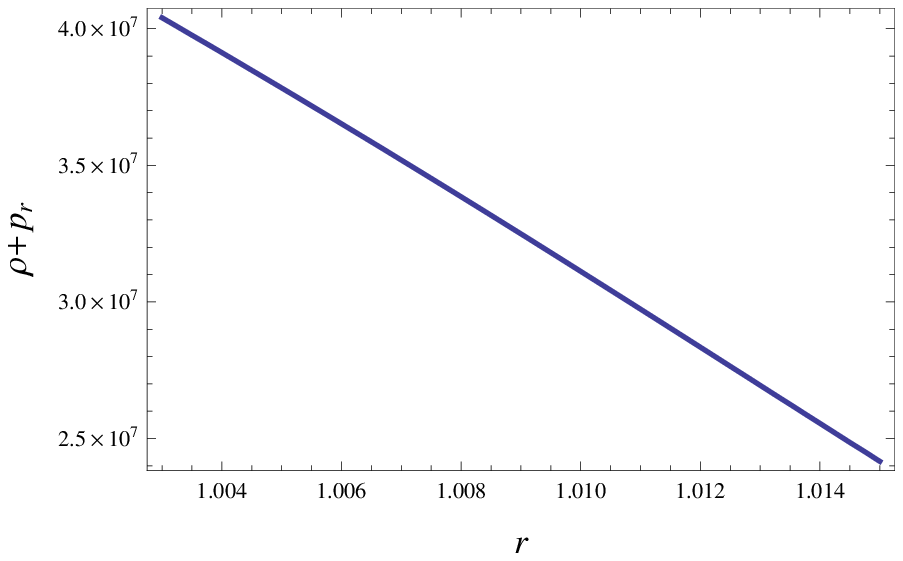,width=0.49\linewidth}
\epsfig{file=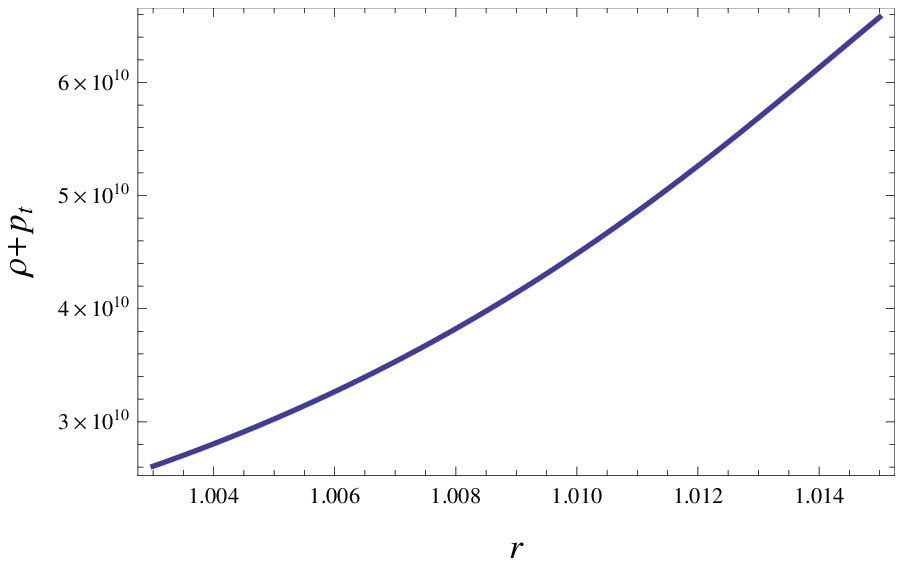,width=0.49\linewidth}
\epsfig{file=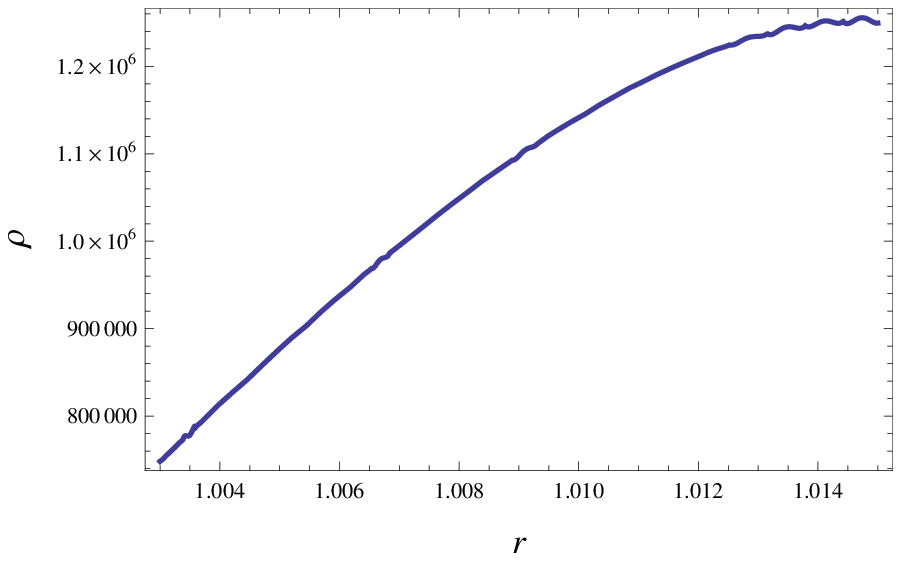,width=0.49\linewidth}\caption{Plots of
$\rho^{(eff)}+p_r^{(eff)}$, $\rho+p_r$, $\rho+p_t$ and $\rho$ versus
$r$ for the first model.}
\end{figure}

As we know that throat radius is the point where $\beta(r)-r$ cuts
the $r$-axis. Here, the throat radius is located at $r_{th}=1.029$
which also satisfies the condition $\beta(r)=r_{th}$ up to two
digits, i.e., $\beta(1.029)=1.028$. The third graph of Figure
\textbf{1} implies that the spacetime does not satisfy the
asymptotically flatness condition. The upper left panel of Figure
\textbf{2} represents the validity of condition (\ref{9pN}). Thus,
the violation of effective NEC confirms the presence of traversable
 wormhole. Also, Figure \textbf{2} shows the plots of $\rho+p_r$ (upper right panel),
$\rho+p_t$ (lower left panel) and $\rho$ (lower right panel) for
normal matter that exhibit positive behavior in the interval
$1.003<r<1.015$. This shows that ordinary matter satisfies the NEC
and physically acceptable wormhole solution is achieved for this
model.

\subsection{Second Model}
\begin{figure}\center
\epsfig{file=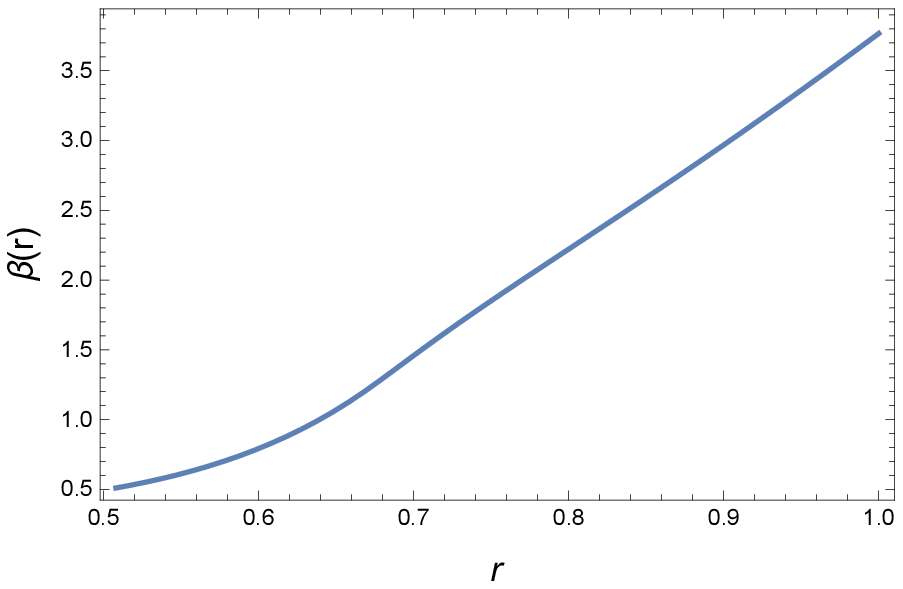,width=0.45\linewidth}
\epsfig{file=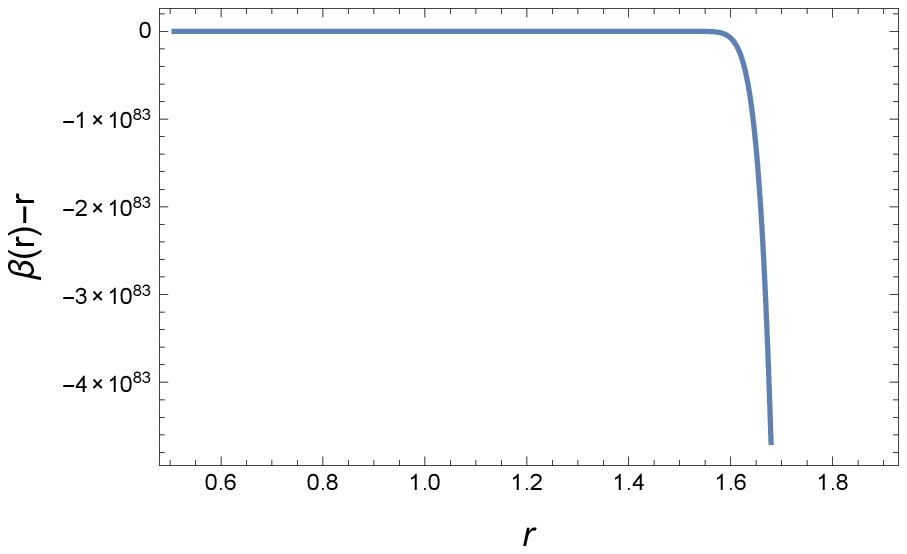,width=0.45\linewidth}
\epsfig{file=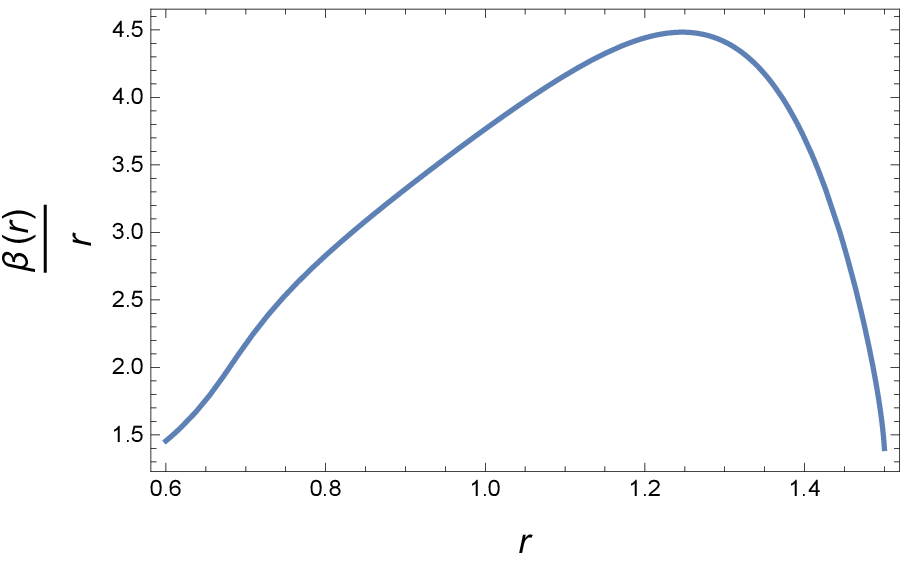,width=0.5\linewidth}\caption{Plots of $\beta(r)$,
$\beta(r)-r$ and $\frac{\beta(r)}{r}$ versus $r$ for the second
model.}
\end{figure}
\begin{figure}\center
\epsfig{file=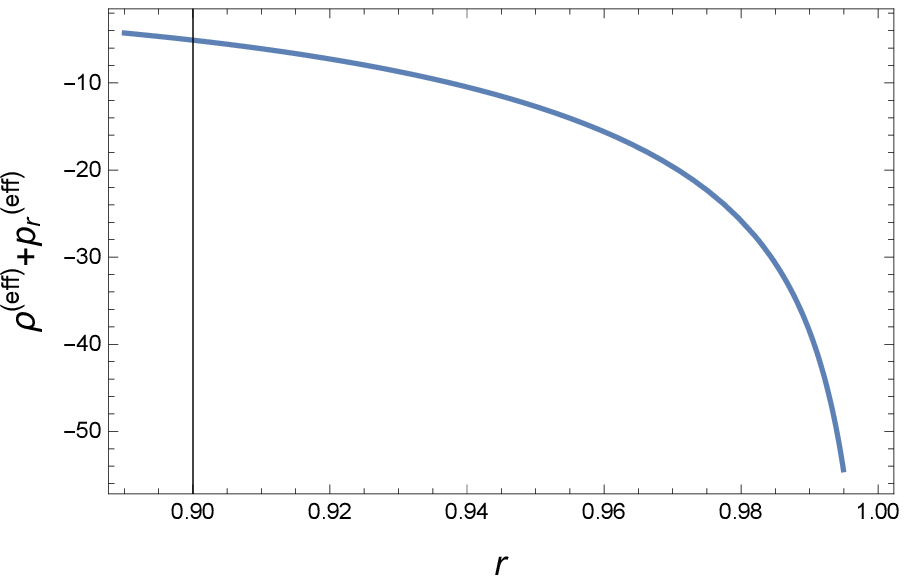,width=0.45\linewidth}
\epsfig{file=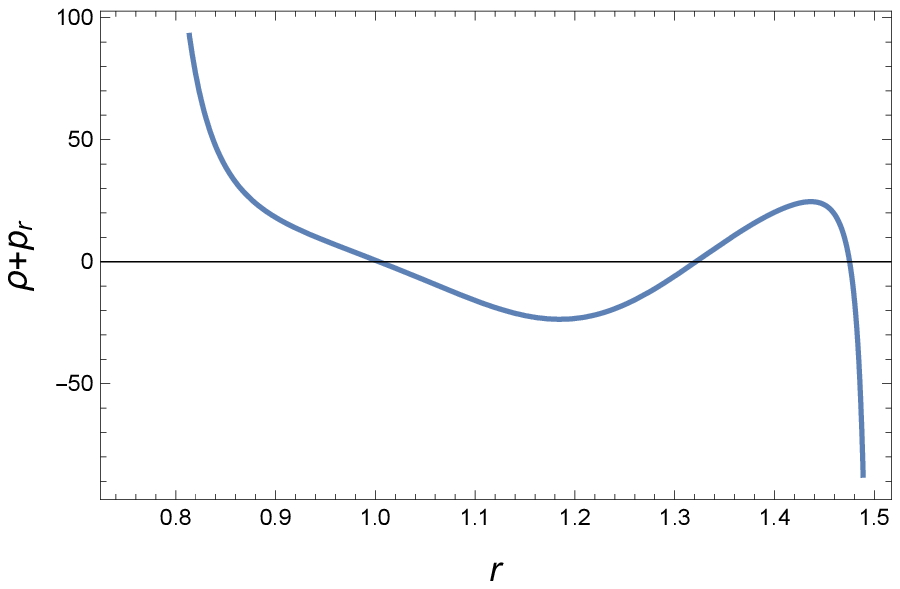,width=0.45\linewidth}
\epsfig{file=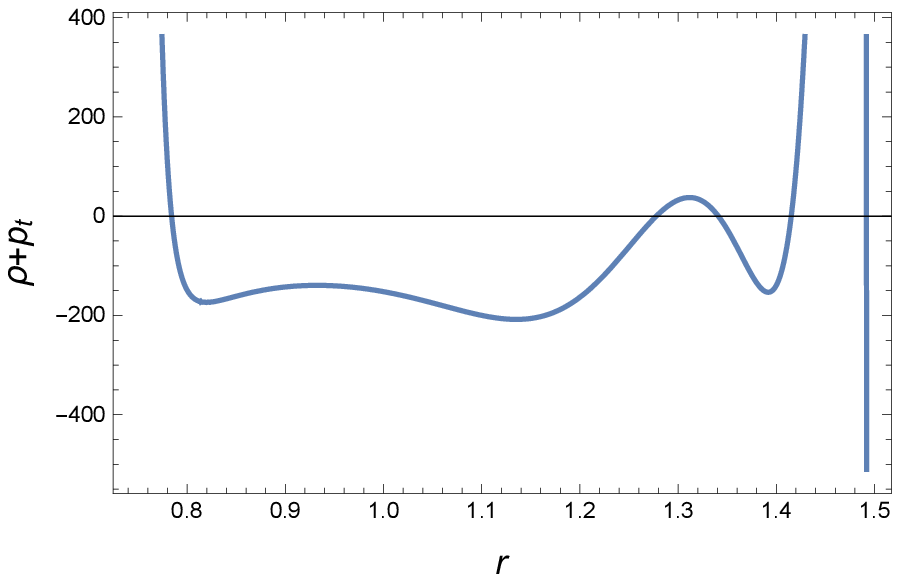,width=0.45\linewidth}
\epsfig{file=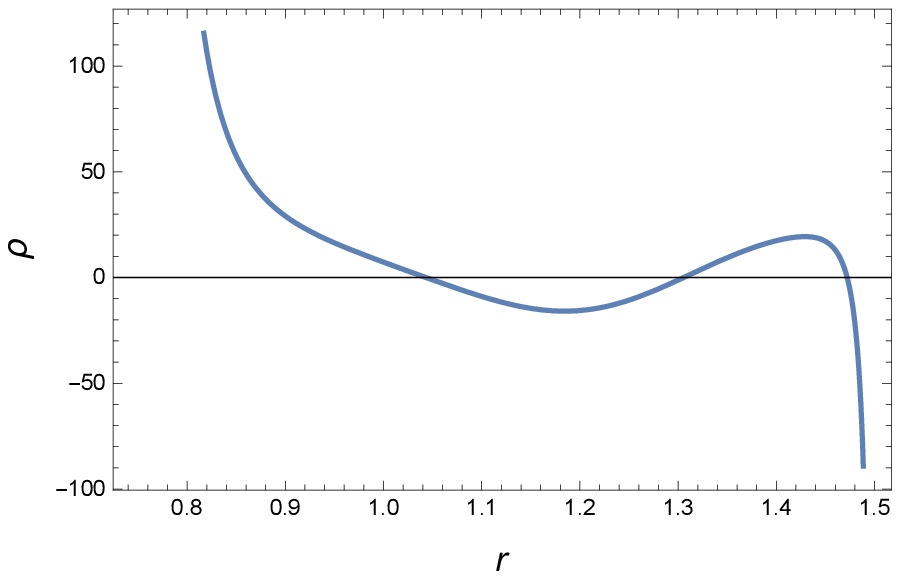,width=0.45\linewidth}\caption{Plots of
$\rho^{(eff)}+p_r^{(eff)}$, $\rho+p_r$, $\rho+p_t$ and $\rho$ versus
$r$ for the second model.}
\end{figure}

We assume the second model as \cite{16}
\begin{equation}\label{13}
F(T,T_G)=-T+\eta_1\sqrt{T^2+\eta_2T_G},
\end{equation}
where $\eta_1$ and $\eta_2$ are the arbitrary constants. We get a
differential equation by substituting Eqs.(\ref{4}), (\ref{5}) and
(\ref{13}) in (12). The numerical technique is used to calculate
$\beta(r)$ from the differential equation by assuming same values of
$\theta$, $\mathcal{M}$ and $\psi$ as above. The model parameters
are taken as $\eta_1=-1.1259$ and $\eta_2=-0.9987$. Also, we take
the following conditions: $\beta(1.5)=2.1$, $\beta'(1.5)=-133.988$
and $\beta''(1.5)=-60000$. We discuss the properties necessary for
the development of wormhole structure. The plot of shape function is
shown in Figure \textbf{3} (left panel) which represents increasing
behavior for all values of $r$. It can be noted that
$\beta(0.5)=0.5$. In the upper right panel, we plot $\beta(r)-r$
versus $r$ to discuss the location of wormhole throat. It can be
observed that small values of $r$ refer as the throat radius.

The lower graph represents the behavior of $\frac{\beta(r)}{r}$. It
can be seen that as the value of $r$ increases, the curve of
$\frac{\beta(r)}{r}$ approaches to $0$. Hence, the spacetime
satisfies asymptotically flatness condition. The upper left panel of
Figure \textbf{4} represents the negative behavior and shows the
validity of condition (\ref{9pN}). For physically acceptable
wormhole solution, we check the graphical behavior of NEC and WEC
for matter energy density and pressure. Figure \textbf{4} shows that
$\rho+p_r$, $\rho+p_t$ and $\rho$ behave positively in the intervals
$1.32\leq r\leq1.474$, $1.28\leq r\leq1.342$ and $1.307\leq
r\leq1.471$, respectively. The common region of these intervals are
$1.28\leq r\leq1.342$. This indicates that NEC and WEC are satisfied
in a very small interval. Thus there can exist a micro or tiny
physically acceptable wormhole for this model. Tiny wormhole means
small radius with narrow throat.

\section{Equilibrium Condition}

In this section, we investigate equilibrium structure of wormhole
solutions. For this purpose, we consider generalized
Tolman-Oppenheimer-Volkov equation in an effective manner as
\begin{equation}\nonumber
-p'^{(eff)}_r-(p^{(eff)}_r+\rho^{(eff)})\left(\frac{\alpha'}{2}\right)
+(p^{(eff)}_t-p^{(eff)}_r)\left(\frac{2}{r}\right)=0,
\end{equation}
with the metric
$ds^2=diag(e^{2\alpha(r)},-e^{\nu(r)},-r^2,-r^2\sin^2\theta)$, where
$e^{\nu(r)}=\left(1\right.$-$\left.\frac{\beta(r)}{r}\right)^{-1}$.
The above equation can be written as
\begin{equation}\label{14L}
-p'^{(eff)}_r-(p^{(eff)}_r+\rho^{(eff)})\left(\frac{M^{(eff)}e^{\frac{\alpha-\nu}{2}}}{r^2}\right)
+(p^{(eff)}_t-p^{(eff)}_r)\left(\frac{2}{r}\right)=0,
\end{equation}
where the effective gravitational mass is described as
$M^{(eff)}=\frac{1}{2}\left(r^2e^{\frac{\nu-\alpha}{2}}\right)\nu'$.
The equilibrium picture describes the stability of corresponding
wormhole solutions with the help of three forces known as
gravitational force $F_{gf}$, anisotropic force $F_{af}$ and
hydrostatic force $F_{hf}$. The gravitational force exists because
of gravitating mass, anisotropic force occurs in the presence of
anisotropic system and hydrostatic force is due to hydrostatic
fluid. We can rewrite Eq.(\ref{14L}) as
\begin{figure}\center
\epsfig{file=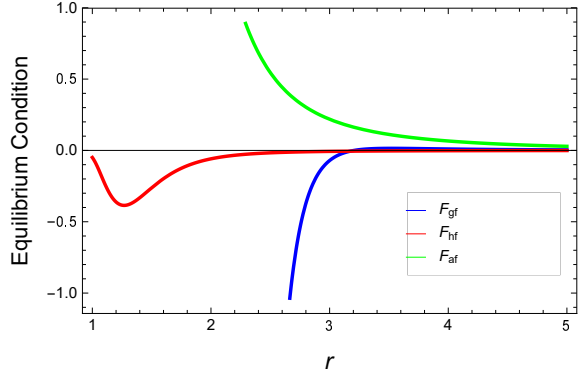,width=0.55\linewidth}\caption{Plot of
equilibrium condition for first model.}
\end{figure}
\begin{figure}\center
\epsfig{file=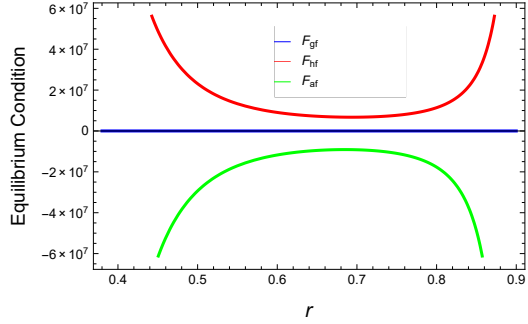,width=0.50\linewidth}\caption{Plot of
equilibrium condition for second model.}
\end{figure}
\begin{equation}\label{16L}
F_{hf}+F_{gf}+F_{af}=0,
\end{equation}
where
\begin{eqnarray}\nonumber
F_{gf}&=&-(p_r^{(eff)}+\rho^{(eff)})\left(\frac{e^{\frac{\alpha-\nu}{2}}M^{(eff)}}{r^2}\right),
F_{hf}=-p'^{(eff)}_r,\\\nonumber
F_{af}&=&(p_t^{(eff)}-p_r^{(eff)})\left(\frac{2}{r}\right).
\end{eqnarray}
Further, we examine the stability of wormhole solutions for first
and second model through equilibrium condition. Using
Eqs.(\ref{7})-(\ref{9}) and (\ref{12}) in (\ref{16L}), we obtain a
difficult equation for the first model. By applying numerical
technique, we plot the graphs of above defined three forces. In
Figure \textbf{5}, it can be easily analyzed that all the three
forces cancel their effects and balance each other in the interval
$4.8\leq r\leq5$. This means that wormhole solution satisfies the
equilibrium condition for the first model. Next, we take the second
model and follow the same procedure by using Eq.(\ref{13}). After
simplification, we finally get a differential equation and solve it
numerically. Figure \textbf{6} indicates that the gravitational
force is zero but anisotropic and hydrostatic forces completely
cancel their effects. Hence, for this model, the system is balanced
which confirms the stability of the corresponding wormhole solution.

\section{Concluding Remarks}

In general relativity, the structure of wormhole is based on the
condition that NEC is violated. This violation supports the fact
that there exist a mysterious matter in the universe famous as
exotic matter and distinguished by its negative energy density. The
amount of this amazing matter would be minimized to obtain a
physically viable wormhole. However, in modified theories, the
situation may be completely different. This paper investigates
noncommutative wormhole solutions with Lorentzian distribution in
$F(T,T_G)$ gravity. For this purpose, we have assumed a diagonal
tetrad and a particular redshift function. We have examined these
wormhole solutions graphically.

For the first model, all the properties are satisfied that are
necessary for wormhole geometry regarding the shape function except
asymptotic flatness. In this case, WEC and NEC for normal matter are
also satisfied. Hence, this model provides realistic wormhole
solution in a small interval threaded by normal matter rather than
exotic matter. The violation of effective NEC confirms the
traversability of the wormhole. Furthermore, the second model
fulfills all properties regarding shape function and also satisfies
WEC and NEC for normal matter. There exists a micro wormhole
solution which is supported by normal matter. This model satisfies
the traversability condition (\ref{9pN}). We have investigated
stability of both models through equilibrium condition. It is
mentioned here that stability is attained for both models.

Bhar and Rahaman \cite{33} examined in GR whether the wormhole
solutions exist in different dimensional noncommutative spacetime
with Lorentzian distribution. They found that wormhole solutions
appear only for four and five dimensions but no solution exists for
higher dimensions. It is interesting to mention here that we have
also obtained wormhole solutions that satisfy all the conditions and
are stable in $F(T,T_G)$ gravity. Our results show consistency with
the teleparallel equivalent of GR limits. For the first model, if we
substitute $\gamma_1=\gamma_3=0$, then the behavior of shape
function $\beta(r)$ and energy conditions in teleparallel theory
remains the same as in this theory. For the second model, $\eta_1=0$
provides no result but if we consider $\eta_2=0$, then $\beta(r)$ as
well as energy conditions represent consistent behavior.

In $F(T)$ gravity \cite{41}, the resulting noncommutative wormhole
solutions are supported by normal matter by assuming diagonal
tetrad. In the underlying work, we have also obtained solutions that
are threaded by normal matter. Kofinas et al. \cite{42} discussed
spherically symmetric solutions in scalar-torsion gravity in which a
scalar field is coupled to torsion with a derivative coupling. They
obtained exact solution which represents a new wormhole-like
solution having interesting physical features. We can conclude that
in $F(T,T_G)$ gravity, noncommutative geometry with Lorentzian
distribution is a more favorable choice to obtain physically
acceptable wormhole solutions rather than noncommutative
geometry \cite{35}.\\\\

\textbf{The authors have no conflict of interest for this research.}

\end{document}